\documentclass[preprint]{jpsj3}
\usepackage{txfonts}

\title{$\Gamma_3$-type Lattice Instability and the Hidden Order of URu$_2$Si$_2$}

\author{Tatsuya \surname{Yanagisawa}$^1$\thanks{E-mail: tatsuya@phys.sci.hokudai.ac.jp}, Shota \surname{Mombetsu}$^1$, Hiroyuki \surname{Hidaka}$^1$,  Hiroshi \surname{Amitsuka}$^1$,\\
 Mitsuhiro \surname{Akatsu}$^{2,3}$, Shadi \surname{Yasin}$^2$, Sergei \surname{Zherlitsyn}$^2$, Jochen \surname{Wosnitza}$^2$,\\
 Kevin \surname{Huang}$^4$, M. Brian \surname{Maple}$^4$}

\inst{
$^1$Department of Physics, Hokkaido University, Sapporo 060-0810, Japan \\
$^2$Hochfeld-Magnetlabor Dresden, Helmholtz-Zentrum Dresden-Rossendorf and TU Dresden, D-01314 Dresden, Germany\\
$^3$Graduate School of Science and Technology, Niigata University, Niigata 950-2181, Japan\\
$^4$Department of Physics, University of California San Diego, La Jolla 92093, USA.
} 

\abst{
We have performed ultrasonic measurements on single-crystalline URu$_2$Si$_2$ with pulsed magnetic fields, in order to check for possible lattice instabilities due to the hybridized state and the hidden-order state of this compound. The elastic constant ($C_{11}$-$C_{12}$)/2, which is associated with a response to the $\Gamma_3$-type symmetry-breaking (orthorhombic) strain field, shows a three-step increase at $H \ge$ 35 T for $H \parallel c$ at low temperatures, where successive meta-magnetic transitions are observed in the magnetization. We discovered a new fact that the absolute change of the softening of ($C_{11}$-$C_{12}$)/2 in the temperature dependence is quantitatively recovered at the suppression of hybridized-electronic state and the hidden order in high-magnetic field for $H \parallel c$ associated with the successive transitions. The present results suggest that the $\Gamma_3$-type lattice instability, is related to both the emergence of the hybridized electronic state and the hidden-order parameter of URu$_2$Si$_2$. On the other hand, magnetic fields $H \parallel$ [100] and [110] enhance the softening of ($C_{11}$-$C_{12}$)/2 in the hidden order phase, while no step-like anomaly is observed up to 68.7 T. We discuss the limitation of the localized-electron picture for describing these features of URu$_2$Si$_2$ by examination of a crystalline electric field model in terms of mean-field theory.
}

\kword{URu$_2$Si$_2$, hidden order, elastic constant, ultrasound, pulsed-magnetic field}

\begin{document}
\maketitle

`What is the primary order parameter and its ordering vector for the hidden order of URu$_2$Si$_2$?' is still an open and controversial question and a longstanding issue of condensed matter physics.~\cite{Palstra85} URu$_2$Si$_2$, which crystallizes in the ThCr$_2$Si$_2$-type tetragonal structure (space group No. 139, I4/mmm), shows an unknown second-order transition at $T_{\rm o}$ = 17.5 K, known as the hidden order (HO) phase, and also a transition into an unconventional superconducting state below $T_{\rm c} \sim$ 1.4 K.~\cite{Palstra85, Maple86, Schlabitz86}
Many theoretical models have been proposed from both localized and itinerant electron pictures in attempts to identify the order parameter of the HO phase.
 Since $^{29}$Si-NMR and zero-magnetic-field $\mu$SR measurements have detected little ($\le$ 1 Gauss) or no significant internal-magnetic field in the HO phase,~\cite{Takagi07, Luke94, Amitsuka02} non-dipolar-type order parameters, {\it e.g.,} electric quadrupole (rank 2), hexadecapole (rank 4), $\Gamma_3$-type magnetic octupole (rank 3), or dotriacontapole (rank 5) order parameters, are recently attracting attention.~\cite{Santini94, Ohkawa99, Haule09, Harima10, Kusunose11, Kiss05, Ikeda12} Thus far, none of the characteristic signatures of the electric-quadrupole or magnetic-octupole order has been identified in resonant X-ray scattering (under magnetic fields) and neutron scattering under uniaxial stress.~\cite{Amitsuka10, Amitsuka13} On the other hand, a `{\it nematicity}' of the electron state in the HO is pointed out, since an in-plane rotational four-fold symmetry breaking is observed in the HO by the magnetic-torque measurement by use of cantilever technique.~\cite{Okazaki11} The interpretation of these experimental results remains controversial.

URu$_2$Si$_2$ is also considered a heavy-fermion compound since several physical properties of this compound exhibit signatures of many-body effects. In particular, the temperature dependence of the electrical resistivity for $I \parallel a$ and the magnetic susceptibility for $H \parallel c$ show maxima at around $T_{\rho,{\rm max}} \sim 70$ K and $T_{\chi,{\rm max}} \sim 55$ K, respectively, and both quantities are strongly suppressed with decreasing temperatures.~\cite{Palstra85, Scheerer12}
The whole feature of the electrical resistivity is similar to the typical Kondo-lattice behavior, whereas the large reduction of the c-axis magnetic susceptibility at low temperature seems to be unconventional, suggesting the influence of crystalline-electric-field (CEF) effects and/or large antiferromagnetic correlations.~\cite{Santini94, Kusunose11, Kiss05, Garatanu05} However, there is no consensus of the mechanism behind the formation of the low-temperature heavy-electron state of this system, even on the CEF 5f level schemes assumed in the local limit.

On the other hand, when applying a magnetic field over 35 T along $c$-axis, three-successive transitions have been observed in resistivity, specific-heat, magnetization, and longitudinal-ultrasonic-velocity measurements.~\cite{Sugiyama90, Wolf01, Jaime02, Suslov03, Scheerer12} These cascade transitions occur in the vicinity of the upper phase boundary of the HO phase and are associated with meta-magnetic-like increases in the $c$-axis magnetization.~\cite{Harrison03, Sugiyama90} Intriguingly, the crossover temperatures of $T_{\rho,{\rm max}}$ and $T_{\chi,{\rm max}}$ are gradually reduced with increasing magnetic field and vanishe in the vicinity of the HO phase boundary in the magnetic-field-temperature phase diagram.~\cite{Scheerer12} This fact suggests that a drastic change from the heavily hybridized to little or non-hybridized  electronic state  is expected with the collapse of the HO. Here, one of the outstanding issues is whether any signature of symmetry breaking can be observed within the cascade-transition region in order to clarify the characteristics of the HO and multiple phases in high magnetic field.

Ultrasonic measurements can sensitively detect symmetry-breaking lattice instabilities and electric quadrupole responses in the single crystals. Early studies employing ultrasonic measurements on URu$_2$Si$_2$ uncovered a characteristic decrease (softening) of the elastic constant $C_{11}$ and ($C_{11}$-$C_{12}$)/2 with decreasing temperature.~\cite{Wolf94, Kuwahara97} The softening of $C_{11}$ had been explained by Gr{\"u}neisen coupling due to many-body effects, which appears in the bulk modulus of typical heavy-fermion compounds. This phenomena is the so-called Kondo-volume collapse.~\cite{Luethi} On the other hand, a study of elastic properties of URu$_2$Si$_2$ in comparison with the non-$5f$ contribution of ThRu$_2$Si$_2$ implies that the softening of ($C_{11}$-$C_{12}$)/2 in URu$_2$Si$_2$ originates from uranium's $5f$ electrons~\cite{Yanagisawa12}. The ($C_{11}$-$C_{12}$)/2 mode, which corresponds to orthorhombic strain that can conserve volume and breaks $\Gamma_3$ symmetry, is the only transverse mode which exhibits softening in this compound. Such a mode-selective elastic response reminds us of a $\Gamma_3$-type crystal deformation due to orbital fluctuations of $c$-$f$ hybridized bands (band Jahn-Teller effect) as an origin of softening rather than a CEF effect with a pseudo degenerate ground state, which potentially causes softening in other transverse modes. They are, however, not fully understood by both itinerant and localized $5f$-electrons pictures, since the $c$-$f$ hybridized band structure and possible contribution of the band for the elastic responses have not been clarified, thus far.

The present study addresses the above issues by means of ultrasonic measurements on single-crystalline URu$_2$Si$_2$ with pulsed-magnetic fields up to 68.7 T at the Hochfeld-Magnetlabor Dresden to check for the elastic response at high magnetic fields. In the present paper, we report on the magnetic-field dependence of the transverse-ultrasonic mode ($C_{11}$-$C_{12}$)/2 for $H \parallel c$ and also for $H \perp c$. This provides new information about the relationship between the HO and symmetry-breaking lattice instability. So far, there are only a few reports concerning ultrasonic studies of the cascade transitions in the high field region using the longitudinal $C_{11}$ and $C_{33}$ modes, which correspond to strain that changes volume while preserving the tetragonal symmetry.~\cite{Wolf01, Suslov03}

The sample used was a single crystal of URu$_2$Si$_2$ with the dimensions of $3.8\times1.8\times1.2$ mm$^3$ ([110]-[1\=10]-[001]). LiNbO$_3$ transducers with thickness of 40 $\mu$m are fixed on the mirror-polished [110] surfaces and used for generation and detection of ultrasound. The sound velocity was measured by use of the phase-comparator method by recording the in-phase and quadrature signals with repetition rate of 20 kHz at fixed frequencies 255.6 MHz and 155.3 MHz for $H \parallel c$ and $H \perp c$, respectively. In the present measurements, an absolute value of the sound velocity for propagation $k \parallel [110] $ and polarization $u \parallel$ [1\=10] was $v =$ 2511 m/s at 4.2 K, which results in an absolute value of the elastic constant ($C_{11}$-$C_{12}$)/2 = 6.453$\times 10^{10}$ J/m$^3$ with the density $\rho = 10.01$ g/cm$^3$ of URu$_2$Si$_2$. The pulsed-magnetic fields had a duration of $\sim$25 ms with maximum field of 46.8 T for $H \parallel c$ and a duration of $\sim$150 ms with maximum field of 68.7 T for $H \perp c$. As shown below, the signal-to-noise ratio of the present measurement is small for all measurements, indicating that the measurement conditions were maintained during the sample rotations. In particular, the sample was manually rotated together with a non-magnetic holder without removing the transducers and signal lines from the sample for the in-plane magnetic field while rotating from $H \parallel$ [100] to [110].

Figure 1 shows the magnetic-field dependence of ($C_{11}$-$C_{12}$)/2 and of the ultrasonic-attenuation coefficient $\Delta \alpha$ at 1.5 K. ($C_{11}$-$C_{12}$)/2 incleases drastically with $H \parallel c$, but decreases moderately for $H \perp c$. This anisotropy shows that the origin of the elastic softening observed in the HO state is being suppressed for $H \parallel c$ while being enhanced for $H \perp c$. For $H \parallel c$, we can see three anomalies in the successive-transition region, 35 T $\le H \le$ 39 T, as indicated by the downward pointing arrows in upper panel of Fig. 1 and inset (a). At 4.2 K, only one kink, which corresponds to the upper HO-phase boundary, can be identified. These transition fields are consistent with previous reports of pulsed- and static-magnetic field measurements~\cite{Harrison03, Scheerer12, Kim03}. This indicates that an {\it isothermal condition} was maintained during the pulse in the present measurement, {\it i.e.}, a magneto-caloric effect (magnetic refrigeration) in the vicinity of the phase boundary is not occurring, while such effects easily occur when the pulse duration is too short. Indeed, the elastic anomalies shift to lower fields and broaden at 4.2 K as shown in inset (a) of Fig. 1. On the other hand, no elastic anomaly is observed for $H \perp c$ up to 68.7 T as shown in inset (b) of Fig. 1. The absolute value of the elastic constant after applying pulsed magnetic fields (at $H$ = 0) always returns to the value observed before the pulse. This shows that there is no eddy-current heating even for the large cross section ($\sim$6.8 mm$^2$) and short pulse-duration for the magnetic field aligned along $c$.
Incidentally, no clear elastic anomaly was detected in the HO phase for $H \parallel c$ at around $H^* \sim 22$ T, where a reconstruction of the Fermi surface is observed from electrical transport measurements using high quality single crystal~\cite{Shishido09, Aoki12}.
A small hysteresis-like loop of uncertain origin is, however, observed in the intermediate magnetic-field region.
 
\begin{figure}[t]
\begin{center}
\vspace{5 mm}
\includegraphics[width=0.9\linewidth]{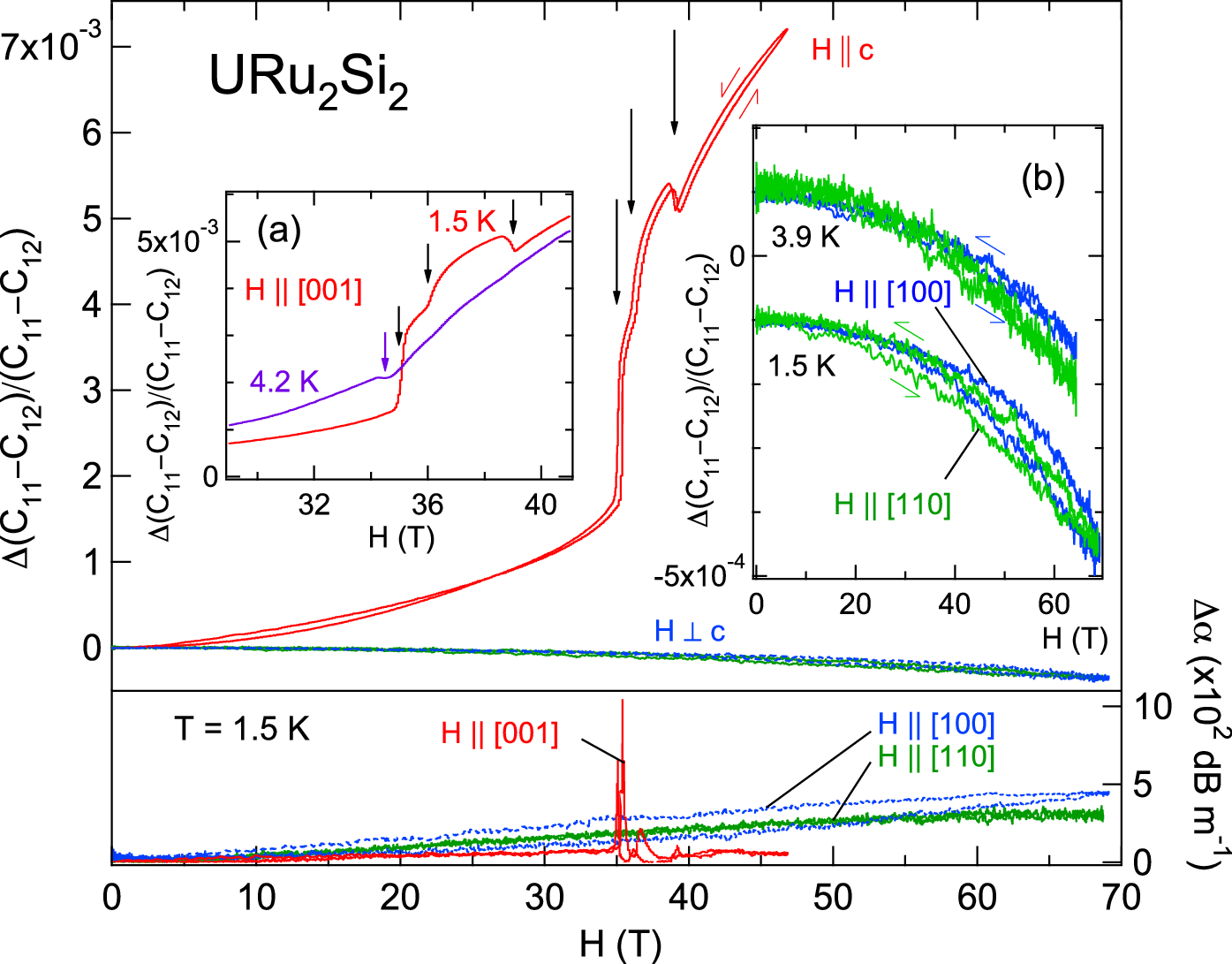}
\end{center}
\caption{(Color online) 
Magnetic-field dependence of elastic constant ($C_{11}$-$C_{12}$)/2 (upper panel) and sound-attenuation change $\Delta \alpha$ (lower panel) for $H \parallel c$ and $H \perp c$ at 1.5 K. Inset (a) shows elastic anomalies in the vicinity of the cascade transitions at 1.5 and 4.2 K. Inset (b) shows ($C_{11}$-$C_{12}$)/2 vs. magnetic field for $H \parallel$ [100] and $H \parallel$ [110] at 1.5 and 3.9 K
The data, except the inset (a), show both up- and down-sweep of the magnetic fields.
}
\label{f1}
\vspace{-5mm}
\end{figure}

The in-plane magnetic field dependence for $H \parallel$ [100] and $H \parallel$ [110] in the HO phase at 1.5 and 3.9 K is shown in inset (b) of Fig. 1. The slight difference of $4 \times 10^{-5}$ in the response of ($C_{11}$-$C_{12}$)/2 can be recognized between the data for $H \parallel$ [100] and [110] in the high magnetic-field region. The difference is, however, within the measurement accuracy of $\sim 5 \times 10^{-5}$ from the noise signal. Thus, we conclude that there is no evidence of in-plane anisotropy. Later, we indeed will show that the expected effect is too small to be observable in our measurement.
\begin{figure}[t]
\begin{center}
\vspace{5mm}
\includegraphics[width=0.9\linewidth]{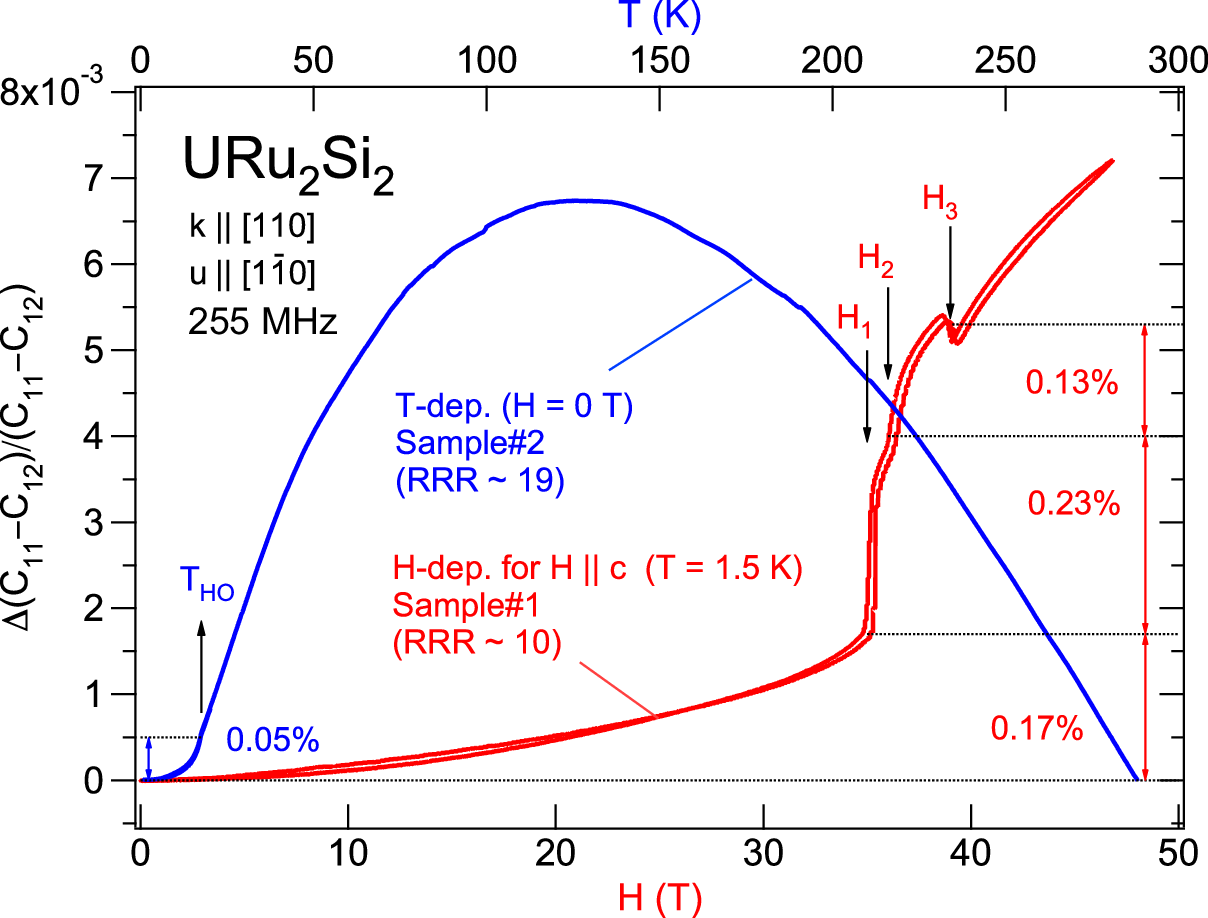}
\end{center}
\caption{(Color online) 
Comparison of elastic constant change in the temperature dependence (blue plots) and magnetic field dependence for $H \parallel c$ (red plots).
}
\label{f2}
\vspace{-5mm}
\end{figure}

In Fig. 2, we present the temperature and magnetic-field dependence of the elastic response for two individual single crystals with similar residual resistivity ratio (RRR = $\rho_{(1.5 K)}/\rho_{(300 K)})$. Sample\#1 has RRR $\sim 10$ and sample\#2 has RRR $\sim 19$. These two data sets are plotted with the same ordinate; the upper axis corresponding to the blue data shows the temperature dependence at 0 T, the lower axis corresponding to the data shows the magnetic-field dependence at 1.5 K. Here, we notice that the elastic-constant change (softening of $\sim$0.7\%) in the temperature dependence below the local maximum at $\sim$120 K is comparable to the increase up to $H \sim$ 45 T in the field dependence. Here, the maximum in the elastic constant eventually occurs as a balance between the phonon-background and the softening of strain-susceptibility due to $5f$-electron effects in the temperature dependence~\cite{Yanagisawa12}, while the $c$-axis magnetization shows a maximum at around $T_{\chi,{\rm max}} \sim$ 55 K.
It should be mentioned that the present temperature dependence is different from the previously reported data~\cite{Kuwahara97}, which shows the maximum at $\sim70$ K and also the softening $\sim0.2$\% for ($C_{11}$-$C_{12}$)/2. We consider these differences are due to the recent improvements of the measurement conditions, {\it e.g.}, using mirror-polished sample, appropriate adhesive agents, and higher frequency (higher directional quality) of the ultrasonic wave.
It should be noted that the $c$-axis magnetization also responds similarly to temperature and magnetic field changes. So we point out that there are considerable similarities between the reduction of $c$-axis magnetization and elastic constant ($C_{11}$-$C_{12}$)/2 in the temperature dependence, and also the different tendencies of the elastic constant between $H \parallel c$ and $H \perp c$ in the magnetic field dependence and the strong anisotropy in the $c$- and $a$-axis magnetization in URu$_2$Si$_2$. These similarities strongly imply that these phenomena will share a common root cause. The origin must be related to the low-temperature electronic state of this compound and also an anisotropic nature of the HO parameter, {\it e.g.}, the higher-rank multipole order or the $c$-$f$ hybridized-band instability.

As described above, since the hybridization effect will be dominant in the low-temperature and low-magnetic field region of this compound, it is unreasonable to reproduce the bulk susceptibilities only based on the localized electron model. Nevertheless, some features can be explained by assuming an appropriate CEF level scheme which mimics the hybridization effect, including the band Jahn-Teller effect, even in the localized $5f$-electron picture.~\cite{Santini94, Kusunose11} In this case, it can be expected that the recovery of the elastic constant ($C_{11}$-$C_{12}$)/2 around the cascade transitions in the high magnetic field region is accompanied by a drastic change of the CEF ground state. In the present paper, we dare to introduce a CEF model in terms of mean-field (MF) theory assuming an antiferro-hexadecapole (AFH) order which reproduce the temperature and magnetic field dependence of ($C_{11}$-$C_{12}$)/2 in both the normal and hybridized states, in order to display the {\it limitations} of the localized-electron picture for describing these responses of ($C_{11}$-$C_{12}$)/2.
\begin{figure}[t]
\begin{center}
\vspace{8mm}
\includegraphics[width=0.9\linewidth]{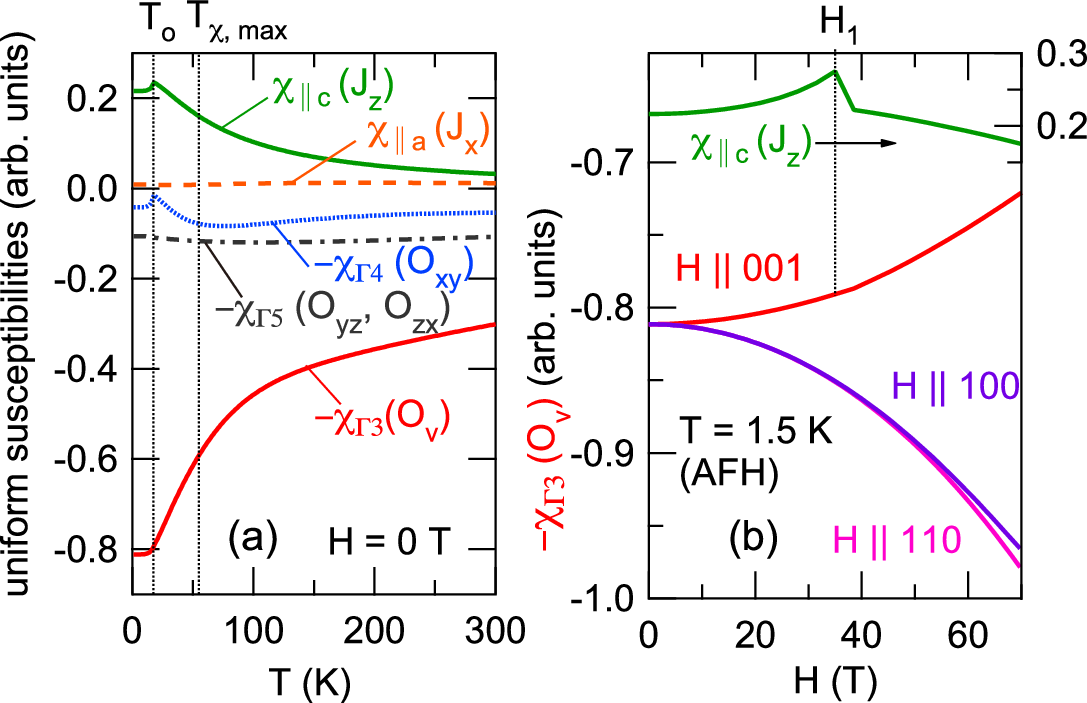}
\end{center}
\caption{(Color online) 
(a): Calculated (uniform) magnetic- and quadrupole-susceptibilities vs. $T$, and (b): calculated (uniform) magnetic- and quadrupole-susceptibilities vs. $H$ for $H \parallel$ [001], [110] and [100] at 1.5 K up to 70 T using the MF theory with a CEF scheme as described in the text.
}
\label{f3}
\vspace{-6mm}
\end{figure}

Figure 3 (a) shows CEF calculations of uniform susceptibilities vs. $T$, and Fig. 3 (b) shows the magnetic-field dependence of these uniform susceptibilities at 1.5 K. These uniform magnetic and quadrupole susceptibilities are obtained by numerical differentiation with the tiny magnetic and strain field, respectively. The present CEF level scheme is; $\Gamma_1^{(1)}$ (0 K)-$\Gamma_2$ (60 K)-$\Gamma_3$ (178 K)-$\Gamma_5^{(1)}$ (491 K), for the Hamiltonian,  $H_{\rm CEF} = -\sum_{m,n} B_m^n O_m^n$ ($l, m = 0, 2, 4, 6$),  with $B_2^0$ = |12 K, $B_4^0$ = |0.43 K, $B_4^4$ = |3.2 K, $B_6^0$ = |0.011 K, and $B_6^4$ = 0.053 K, and the MF parameters (as will be described later). Note that the present CEF model~\cite{Kusunose12} has a similar level scheme as originally proposed for PrRu$_2$Si$_2$.~\cite{Michalski00}
Based on the localized-electron picture, the temperature and magnetic-field dependence of ($C_{11}$-$C_{12}$)/2 is understood as a quadrupole susceptibility $-\chi_{\Gamma 3}$ with the $\Gamma_3$-symmetry quadrupole operator $O_v = \frac{2}{\sqrt{3}}O_2^2$, which is defined as $O_v = (J_+^2+J_-^2)/2$. The present CEF calculation can reproduce the following experimental facts; an anisotropy between $c$- and $a$-axis magnetizations at high temperature region $T \geq$ 100 K, a softening of ($C_{11}$-$C_{12}$)/2 in the temperature dependence, and also an anisotropic response of ($C_{11}$-$C_{12}$)/2 and magnetization for $H \parallel c$ and $H \perp c$. Needless to say, other CEF models, which have been proposed thus far, can not reproduce the anisotropic elastic response, {\it e.g.,} Santini's 5$f^2$ model~\cite{Santini94}, $\Gamma_4$-$\Gamma_1^{(1)}$ (44 K)-$\Gamma_2$ (111 K), and Garatanu's 5$f^2$ model~\cite{Garatanu05}, $\Gamma_5^{(1)}$-$\Gamma_1^{(1)}$ (404 K)-$\Gamma_2$ (1076 K), give rise to softening for $C_{44}$ or Nieuwenhuys's model~\cite{Kusunose11}, $\Gamma_1^{(1)}$-$\Gamma_2$ (50 K)-$\Gamma_1^{(2)}$ (171 K),  does not even reproduce the softening of ($C_{11}$-$C_{12}$)/2.~\cite{Yanagisawa12}

Regarding the possibilities of the AFH order scenario on the localized-electron picture, which has recently been proposed by Kusunose and Harima~\cite{Kusunose11}, the present CEF scheme, which includes a $\Gamma_{\rm 2g}$-symmetry $xy(x^2-y^2)$-type hexadecapole $H_z^{\alpha}$ as an active multipole, does not forbid the AFH order. By using the mean-field (MF) theory of the $H_z^{\alpha}$-type AFH order with the CEF model, the effective Hamiltonian is written as, $H_{\rm MF}^{\rm eff} = \frac{1}{2z}\sum_{(i,j)}^{m.n.}[D_h H_z^{\alpha}(i)H_z^{\alpha}(j)+ D_d J_z(i)J_z(j)]+\sum_{i} H_{\rm CEF}-g_J \mu_{\rm B} H\sum_{i} J_z(i)$. Here, $D_h$ and $D_d$ are coupling constants for the inter-site electric hexadecapole and magnetic dipole interactions, respectively. The additional depression of ($C_{11}$-$C_{12}$)/2 just below $T_{\rm o}$ in the temperature dependence can also be reproduced by the quadrupole susceptibility $-\chi_{\Gamma 3}$ assuming the MF parameters as, $D_d$ = 2.093 K, $D_h$ = 0.0433 K. On the other hand, the present MF model can not reproduce a temperature dependence of the elastic constant $C_{66}$, where the calculated quadrupole susceptibility of $O_{xy} = (J_+^2-J_-^2)/2i$ decreases with decreasing temperature below $T_{\rm o}$ while the experimental result of $C_{66}$ exhibits an opposite temperature dependence in the HO phase.~\cite{Wolf94, Kuwahara97} Moreover, the magnetic-field dependence of $-\chi_{\Gamma 3}$ for  $H \parallel$ [001]  does not show any anomaly at the suppression of the HO, while the calculated $c$-axis magnetic susceptibility shows a jump at $H_1$.
These discrepancies simply show the limitations of the present localized-electron model to describe the elastic responses.

After all, it is undeniable whether the softening of $C_{11}$ is caused by the change of bulk modulus $C_{\rm B}$ due to Kondo effect or a softening of ($C_{11}$-$C_{12}$)/2, since the $C_{11}$ mode in the tetragonal symmetry includes both contributions, {\it i.e.,} it is decomposed as $C_{11}$ = 3$C_{\rm B}$-$C_u$+($C_{11}$-$C_{12}$)/2-2$C_{13}$ with bulk modulus $C_B = (2C_{11}+2C_{12}+4C_{13}+C_{33})/9$ and tetragonal strain mode $C_u = (C_{11}+C_{12}-4C_{13}+2C_{33})/6$. In order to estimate the contribution of $C_{\rm B}$ on $C_{11}$, further ultrasonic measurements will be needed to check the longitudinal mode, with $k \parallel u \parallel [110]$, $C_{\rm L110}$ = 3$C_{\rm B}$-$C_u$+$C_{66}$-2$C_{13}$, and the other transverse modes $C_{44}$ and $C_{66}$ for comparison.

Finally, we will focus on the in-plane magnetic-field dependence at 1.5 K, where no phase transition is found up to 68.7 T (Fig. 3(b)). The difference of $-\chi_{\Gamma 3}$ between $H \parallel$ [100] and [110] in the calculated results in the high magnetic field region is mainly attributable to the CEF effect with or without MF interactions. Despite this, it can be expected that the bulk susceptibility must be affected by induced lower-rank multipoles when applying high magnetic fields $H \perp c$ in the AFH ordered phase. For example, a magnetic dipole and an electric quadrupole should be induced via Ginzburg-Landau coupling.~\cite{Kusunose12}
When the change of calculated susceptibility is scaled to the actual elastic constant change from 0 T to 60 T, the contribution of the difference between $H \parallel$ [100] and [110] is estimated as $\sim 0.5 \times 10^{-5}$ at 60 T. The magnitude of the modulation due to MF interactions accounts for less than 1 \% of this small anisotropy. As described above, even the possible difference between $H \parallel$ [100] and [110] due to the CEF effect could not be distinguished within the present margin of the error $\sim 5 \times 10^{-5}$. 

In conclusion, we have performed, for the first time, the ultrasonic measurement of ($C_{11}$-$C_{12}$)/2 in the HO phase of URu$_2$Si$_2$ for $H \parallel c$ and $H \perp c$ under pulsed magnetic fields up to 68.7 T.  Elastic anomalies of ($C_{11}$-$C_{12}$)/2, which is a symmetry-breaking (volume-conserving) ultrasonic mode, is observed at around the cascade transition region 35 T $\le H \le$ 39 T. For $H \perp c$, we found that there is no clear difference between $H \parallel$ [100] and [110] within the present measurement accuracy of $\sim 5 \times10^{-5}$. The present experimental results suggest that the suppression of the HO in high magnetic field is accompanied by not only the recovery from the heavy-fermion state, which causes a reduction of the $c$-axis magnetic moment at low temperature, but also the reduction of the $\Gamma_3$ lattice instability. We conclude that the low-temperature electronic state of URu$_2$Si$_2$ has a $\Gamma_3$-type ($x^2$-$y^2$-type) lattice instability which is related to the origin of the hybridized electronic state and the HO, such as higher-multipole fluctuations and/or band Jahn-Teller effects with a $\Gamma_3$ symmetry. Thus, an understanding of the 55-120 K energy-scale phenomena due to strong hybridization based on the itinerant electron pictures, as seen in the magnetization and elastic constant ($C_{11}$-$C_{12}$)/2, should be refocused to elucidating the true nature of the HO.

The authors would like to thank Prof. Hiroaki Kusunose for his helpful advice with the mean-field calculations. This work was supported by JSPS KAKENHI Grant No. 23740250 and No. 23102701, EuroMagNET II under the EU contract No. 228043, and U.S. DOE. Grant No. DE-FG02-04-ER46105. One of the authors (M.A) was supported by the Strategic Young Researcher Overseas Visits Program for Accelerating Brain Circulation from the JSPS.

\end{document}